\documentclass{article}

\usepackage{arxiv}
\usepackage[numbers,sort&compress]{natbib}
\usepackage{tabularray}
\usepackage{adjustbox}
 \usepackage{amsmath}
\usepackage[utf8]{inputenc} 
\usepackage[T1]{fontenc}    
\usepackage{hyperref}       
\usepackage{url}            
\usepackage{booktabs}       
\usepackage{amsfonts}       
\usepackage{nicefrac}       
\usepackage{microtype}      
\usepackage{lipsum}		
\usepackage{graphicx}
\usepackage{natbib}
\usepackage{doi}

\usepackage{amsmath}
\usepackage{graphicx}
\usepackage{booktabs}
\usepackage{array}
\usepackage{tabularray}
\usepackage{adjustbox}
\usepackage{booktabs, multirow}
\usepackage{booktabs}
\usepackage{threeparttable}

\title{DiffSSMEye: Conditional High-Frequency Eye Movement Synthesis with Structured State-Space Diffusion}
\renewcommand{\headeright}{}
\renewcommand{\undertitle}{}
\renewcommand{\shorttitle}{}

\date{} 					

\author{
Kamrul Hasan, Mehedi Hasan Raju, Oleg V. Komogortsev\\
Texas State University, San Marcos, Texas, USA\\
{\tt\small \{kamrul.hasan, m.raju, ok\}@txstate.edu}
}

\renewcommand{\shorttitle}{DiffSSMEye: Conditional High-Frequency Eye Movement Synthesis with Structured State-Space Diffusion}

\hypersetup{
pdftitle={A template for the arxiv style},
pdfsubject={q-bio.NC, q-bio.QM},
pdfauthor={David S.~Hippocampus, Elias D.~Striatum},
pdfkeywords={First keyword, Second keyword, More},
}

\begin{document}
\maketitle

\begin{abstract}
Synthesizing high-frequency eye movements requires preserving long-range temporal dependencies and subject-specific oculomotor dynamics. However, methods based on generative adversarial networks (GANs) and variational autoencoders (VAEs) can struggle to preserve these characteristics simultaneously. We introduce DiffSSMEye, a conditional denoising diffusion probabilistic model with bidirectional Structured State Space Sequence (S4) layers for generating five-second gaze-velocity sequences at 1,000 Hz. Generation is conditioned on a coarse gaze reference, user identity, and task, where identity and task embeddings are incorporated through feature-wise linear modulation. Identity-preservation, multi-resolution spectral, and integrated-drift losses complement noise prediction to constrain biometric consistency, frequency structure, and displacement error. Subject-disjoint evaluation on GazeBase across seven tasks yields mean real--synthetic embedding cosine similarities of 0.91--0.95. DiffSSMEye also achieves the lowest median spatial errors among these baselines for horizontal and random saccade tasks. These results demonstrate improved spatial accuracy and biometric feature consistency for subject-specific gaze synthesis.
\end{abstract}


\section{Introduction}
Eye-tracking data provide a rich behavioral representation of human visual attention, cognitive processing, and task-dependent user behavior \cite{rayner2009eye}. These signals support attention-aware interfaces and adaptive human-computer interaction \cite{hutt2021breaking,menges2019improving}, while their subject-specific characteristics enable eye-movement biometrics for user authentication \cite{lohr2022ekyt}. Eye tracking also supports gaze-contingent rendering in extended reality (XR) environments and driver monitoring \cite{patney2016towards, singh2011eye}. In clinical research, eye-movement patterns have been investigated for the computational assessment of dyslexia, autism spectrum disorder (ASD), and attention-deficit/hyperactivity disorder (ADHD) \cite{haller2022eye,thanarajan2023eye,merzon2022eye}. These applications motivate computational methods that capture both task-dependent gaze behavior and individual oculomotor characteristics.

High temporal resolution is particularly important for gaze-based applications that depend on rapid saccadic dynamics, fixational micro-movements, and biometric information encoded in fine-grained velocity patterns \cite{lohr2022ekyt}. Research-grade systems, including the EyeLink 1000 Plus \cite{eyelink1000plus} and Tobii Pro Spectrum \cite{tobiiprospectrum}, support sampling rates of up to 2,000 Hz and 1,200 Hz, respectively. However, acquiring high-quality data requires proper calibration, participant cooperation, and substantial investment in equipment, personnel, and recording time. Even with publicly available resources such as GazeBase \cite{griffith2021gazebase}, collecting similar diverse recordings across individuals, tasks, and longitudinal sessions remains challenging. Furthermore, gaze data can expose identity and sensitive personal attributes, complicating data sharing and reuse \cite{steil2019privacy}. These challenges motivate synthetic eye-movement generation as a complementary source of data for developing and evaluating gaze-based systems.

Synthetic eye-movement data should reproduce the spatial, temporal, and subject-specific characteristics relevant to the intended application. Such data can supplement limited recordings and support controlled experimentation without requiring additional acquisition for every generated sequence. However, synthetic generation does not inherently guarantee privacy, particularly when preserving users' identity is an explicit objective. Deep generative approaches, including generative adversarial networks (GANs), variational autoencoders (VAEs), and denoising diffusion probabilistic models (DDPMs), provide data-driven mechanisms for learning complex signal distributions \cite{fuhl2021fully,prasse2023sp,ho2020denoising}. For example, SP-EyeGAN generates fixation and saccade sequences to construct synthetic eye movements \cite{prasse2023sp}, while VAE-based approaches learn latent representations for generating raw eye-tracking data \cite{fuhl2021fully}. DiffEyeSyn subsequently demonstrated user-specific, high-frequency synthesis through a conditional DDPM \cite{jiao2024diffeyesyn}. These advances establish the feasibility of learned gaze synthesis, while motivating further investigation of temporal modeling and signal-level constraints. In particular, biometric embedding alignment alone does not explicitly constrain spectral structure or the displacement accumulated when generated velocities are integrated into gaze positions.

To address these modeling objectives, we introduce DiffSSMEye, a conditional DDPM whose denoiser incorporates bidirectional Structured State Space Sequence (S4) layers \cite{gu2022s4}. Building on structured state-space diffusion architectures \cite{alcaraz2023sssd,alcaraz2023ecg}, the framework is adapted to generate five-second, two-dimensional gaze-velocity sequences sampled at 1,000~Hz. A coarse reference trajectory provides temporal guidance, while pretrained Eye Know You Too (EKYT) identity embeddings and learned task embeddings condition generation through feature-wise linear modulation \cite{lohr2022ekyt, perez2018film}. Bidirectional S4 layers allow the denoiser to incorporate preceding and subsequent context throughout each segment. To complement the diffusion noise-prediction objective, we combine an identity-preservation loss with multi-resolution spectral and integrated-drift losses. The spectral objective matches short-time Fourier magnitude representations across temporal resolutions, whereas the drift objective penalizes final displacement error after velocity integration. 
Together, these components are designed to model individual gaze dynamics while constraining frequency structure and accumulated positional error. 
The main contributions of this work are as follows:

\begin{enumerate}
    \item We propose DiffSSMEye, a conditional diffusion framework that integrates bidirectional S4 denoising with coarse-reference, identity, and task conditioning for high-frequency, subject-specific eye-movement synthesis.

    \item We combine identity with multi-resolution spectral and integrated-drift losses to constrain embedding consistency, frequency-domain structure, and displacement error during diffusion training.

    \item We conduct a subject-disjoint evaluation to assess spatial accuracy and precision, embedding similarity, and qualitative signal fidelity. The results show the lowest median target-alignment errors among the evaluated generators and the highest or tied-highest embedding similarity across tasks.
\end{enumerate}

\section{Related Work}

\subsection{Statistical Models}
Early approaches to eye movement synthesis employed statistical, procedural, and physiologically motivated models, drawing on signal processing and computer graphics \cite{lee2002eyes, duchowski2015modeling, duchowski2016eye}. 
For example, \citet{lee2002eyes} synthesized eye movements for animated faces using empirical saccade characteristics and statistical patterns derived from human eye-tracking data. Subsequently, \citet{ma2009natural} modeled statistical dependencies between head motion and gaze direction to generate coordinated eye movements from a given head trajectory. 
These studies advanced the behavioral and visual realism of synthetic eyes, although their objectives span distinct problems, including character animation, multimodal motion synthesis, and appearance-based gaze estimation.

Other approaches incorporated explicit oculomotor dynamics into the simulation of gaze signals. Parameterized simulators modeled saccadic velocity profiles and smooth-pursuit dynamics, supporting the generation of eye movement sequences at configurable sampling rates \cite{fuhl2018eye_Kasneci, fuhl2018eye_Santini}. \citet{yeo2012eyecatch} introduced EyeCatch, a visuomotor framework that uses Kalman filtering \cite{welch1995introduction} to estimate target motion and guide coordinated gaze and body movements during object interception. \citet{campbell2014saliency} proposed a unified Bayesian framework for jointly modeling scanpath dynamics and visual saliency.
More recently, \citet{lan2022eyesyn} introduced EyeSyn, a training-free synthesis framework that uses psychologically motivated models to generate fixations, saccades, and smooth pursuits. 
Overall, these approaches offer interpretable control over movement parameters and can incorporate behavioral variability and measurement noise. However, configurable variability does not, by itself, establish that synthetic trajectories preserve the biometric characteristics of a particular individual. 

\subsection{Machine-learning Models}
Machine-learning approaches learn gaze distributions and temporal dependencies directly from recorded eye movements. Early studies \cite{simon2016automatic, assens2018pathgan, goodfellow2020generative} combined visual feature extraction with sequential modeling to predict image-conditioned scanpaths. For example, \citet{simon2016automatic} employed a convolutional neural network (CNN) with long short-term memory (LSTM) modules to generate scanpaths from static images. Similarly, \citet{assens2018pathgan} introduced PathGAN, which combines image features and recurrent networks within an adversarial learning framework to predict fixation sequences. These approaches model the spatial association and sequential structure of visual attention. However, fixation-level predictions do not directly represent the densely sampled trajectories and velocity profiles occurring within and between gaze events.

Beyond image-conditioned prediction, researchers have investigated the generation of raw gaze signals \cite{fuhl2021fully, de2022next, fuhl2022hpcgen}. \citet{fuhl2021fully} employed fully convolutional networks for eye-tracking data segmentation and reconstruction, together with a variational autoencoder (VAE) for generation. 
Subsequently, SP-EyeGAN \cite{prasse2023sp} introduced separate generative adversarial networks for fixation and saccade synthesis, assembling the generated events into continuous gaze sequences according to supplied fixation locations. Although evaluated using reading data, its generation procedure does not explicitly condition on a target user's biometric identity. DiffGaze \cite{jiao2025diffgaze} extended diffusion-based generation to continuous gaze trajectories conditioned on $360^{\circ}$ images and evaluated synthesis at 30~Hz. Its objective emphasizes stimulus-dependent gaze behavior rather than high-frequency, identity-conditioned synthesis.

Addressing user-specific generation, \citet{jiao2024diffeyesyn} introduced DiffEyeSyn, a conditional diffusion framework built on the DiffWave architecture \cite{kong2020diffwave}. DiffEyeSyn conditions synthesis on a coarse reference sequence and biometric embeddings extracted by a pre-trained gaze-authentication model. 
Subsequently, \citet{hasan2025quantitative} compared subject-aware diffusion and GAN models, together with an auxiliary VAE baseline, reporting stronger real--synthetic biometric embedding similarity for the diffusion-based approach. These findings support explicitly incorporating identity information into high-frequency gaze synthesis.

Alongside identity conditioning, denoiser design determines how temporal context is integrated during synthesis. DiffWave uses bidirectional dilated convolutions whose per-pass receptive field depends on network depth, kernel size, and dilation schedule \cite{kong2020diffwave}. Although successive diffusion steps can extend information propagation, structured state-space models (SSMs) offer an alternative based on learned latent dynamics. Structured State Space Sequence (S4) layers admit an equivalent convolutional representation with sequence-spanning kernels, enabling efficient parallel computation \cite{gu2022s4}. Bidirectional S4 processing integrates context from both temporal directions to model local gaze dynamics and dependencies between distant events. Previous studies have incorporated SSMs into diffusion models for time-series imputation, forecasting, and conditional physiological signal generation \cite{alcaraz2023sssd,alcaraz2023ecg}. Building on this work, DiffSSMEye combines a bidirectional S4 denoiser with user, task, and coarse-reference conditioning within a conditional DDPM, using auxiliary objectives to constrain biometric consistency, spectral structure, and integrated displacement.

\section{Proposed Architecture}
\label{sec:proposed_architecture}

\subsection{Overview}
\label{sec:architecture_overview}

\begin{figure*}[t]
    \centering
    \includegraphics[width=0.99\textwidth]{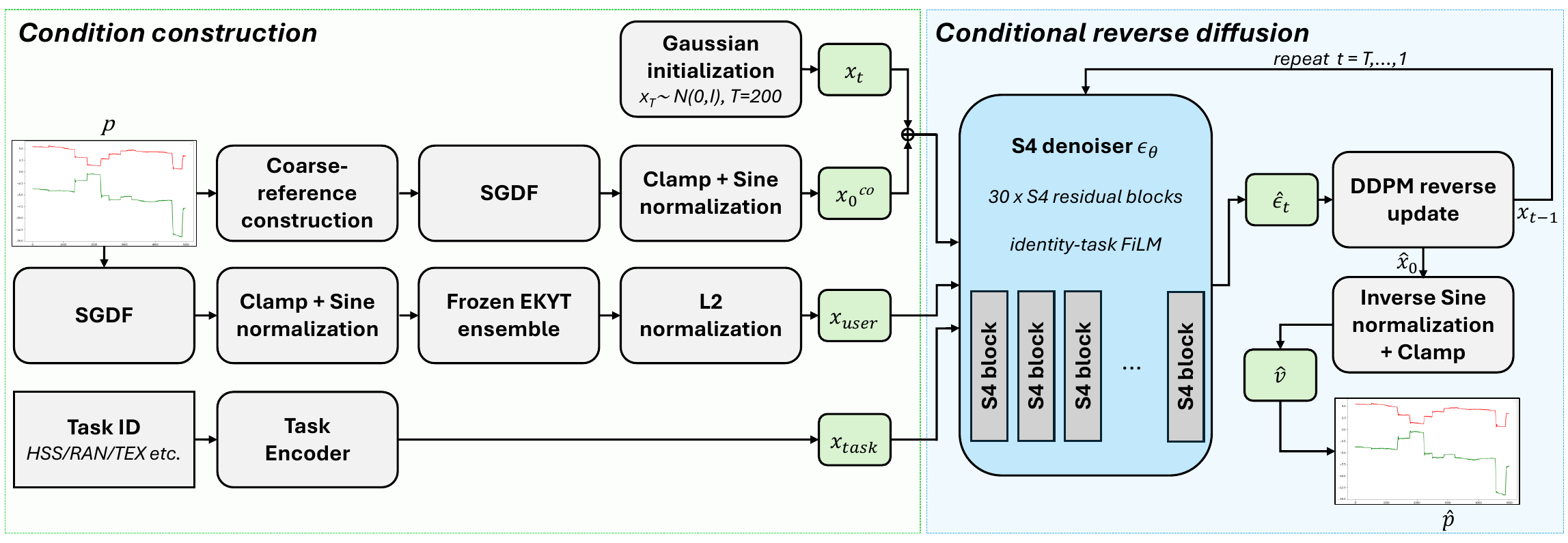}
    \caption{Overview of DiffSSMEye: a DDPM with bidirectional S4 residual blocks transforms Gaussian noise into gaze velocity, conditioned on a coarse reference, an EKYT identity embedding, and a task embedding.}
    \label{fig:main_architecture}
\end{figure*}

DiffSSMEye is a conditional DDPM that synthesizes five-second,
two-dimensional gaze-velocity sequences at $f_s=1000$~Hz ($S=5000$).
Its denoiser utilizes bidirectional S4 layers \cite{gu2022s4}.
As shown in Figure~\ref{fig:main_architecture}, noise prediction uses
the noisy normalized velocity $\mathbf{x}_t$, diffusion step $t$,
coarse reference $\mathbf{x}_0^{\mathrm{co}}$, identity embedding
$\mathbf{e}_{\mathrm{user}}$ from a frozen EKYT encoder, and learned
task embedding $\mathbf{e}_{\mathrm{task}}$:
\begin{equation}
\widehat{\boldsymbol{\epsilon}}_t
=
\boldsymbol{\epsilon}_{\theta}
\left(
\mathbf{x}_t,t,\mathbf{x}_0^{\mathrm{co}},
\mathbf{e}_{\mathrm{user}},\mathbf{e}_{\mathrm{task}}
\right).
\label{eq:conditional_noise_predictor}
\end{equation}
At inference, iterative denoising transforms
$\mathbf{x}_T\sim\mathcal{N}(\mathbf{0},\mathbf{I})$ into
$\widehat{\mathbf{x}}_0$, which is denormalized and integrated to
recover gaze position.

\subsection{Pre-processing}
\label{sec:preprocessing}

Let $\mathbf{p}\in\mathbb{R}^{S\times2}$ be a horizontal--vertical gaze
position segment. 
Initially, a coarse reference is constructed by sampling $\mathbf{p}$ on a $25$-Hz grid and linearly interpolating it back to $1000$~Hz (i.e., $\mathbf{p}^{\mathrm{co}}$). This operation preserves low-frequency trajectory structure without any explicit identity features \cite{raju2021determining}.


The original and coarse positions are differentiated with a
Savitzky--Golay differentiation filter (SGDF) \cite{savitzky1964smoothing} using window length
$7$, polynomial order $2$, and derivative order $1$. After replacing
non-finite values by zero, velocity is clipped to
$v_{\max}=1000$ degrees/s and mapped to $[-1,1]$:
\begin{equation}
\begin{aligned}
\mathbf{v}
&=
\operatorname{clip}
\left(
f_s\operatorname{SGDF}^{(1)}_{7,2}(\mathbf{p}),
-v_{\max},v_{\max}
\right),\\
\mathbf{v}^{\mathrm{co}}
&=
\operatorname{clip}
\left(
f_s\operatorname{SGDF}^{(1)}_{7,2}(\mathbf{p}^{\mathrm{co}}),
-v_{\max},v_{\max}
\right),\\
\mathbf{x}_0
&=
\sin\left(\frac{\pi\mathbf{v}^{\top}}{2v_{\max}}\right),
\qquad
\mathbf{x}_0^{\mathrm{co}}
=
\sin\left(\frac{\pi(\mathbf{v}^{\mathrm{co}})^{\top}}{2v_{\max}}\right).
\end{aligned}
\label{eq:velocity_preprocessing}
\end{equation}


Later, four frozen 128-dimensional EKYT encoders \cite{lohr2022ekyt} extract identity from the original velocity. Their outputs are averaged and $\ell_2$-normalized, while all the tasks are represented by a learned 128-dimensional embedding:
\begin{equation}
\overline{\mathbf{e}}_{\mathrm{user}}
=
\frac{1}{4}\sum_{k=1}^{4}\phi_k(\mathbf{v}),
\qquad
\mathbf{e}_{\mathrm{user}}
=
\frac{\overline{\mathbf{e}}_{\mathrm{user}}}
{\|\overline{\mathbf{e}}_{\mathrm{user}}\|_2+\varepsilon_e},
\qquad
\mathbf{e}_{\mathrm{task}}
=
\operatorname{Emb}_{\mathrm{task}}(r).
\label{eq:mean_ekyt_embedding}
\end{equation}
where $r$ denotes the task label and $\varepsilon_e=10^{-8}$ provides numerical stability.

\subsection{Conditional DDPM}
\label{sec:conditional_ddpm}

The forward process corrupts the clean normalized velocity signal using:
\begin{equation}
\mathbf{x}_t
=
\sqrt{\overline{\alpha}_t}\mathbf{x}_0
+
\sqrt{1-\overline{\alpha}_t}\boldsymbol{\epsilon},
\qquad
\boldsymbol{\epsilon}\sim\mathcal{N}(\mathbf{0},\mathbf{I}),
\qquad
\alpha_t=1-\beta_t,\quad
\overline{\alpha}_t=\prod_{j=1}^{t}\alpha_j.
\label{eq:forward_diffusion}
\end{equation}
We use $T=200$ and a linear schedule as follows:
\begin{equation}
\beta_t
=
\beta_{\mathrm{start}}
+
\frac{t-1}{T-1}
\left(
\beta_{\mathrm{end}}-\beta_{\mathrm{start}}
\right),
\qquad
\beta_{\mathrm{start}}=10^{-4},
\quad
\beta_{\mathrm{end}}=0.02.
\label{eq:linear_beta_schedule}
\end{equation}
For this schedule, $\overline{\alpha}_T\approx0.13218$.

The reverse distribution is parameterized by the conditional noise prediction in Equation~\eqref{eq:conditional_noise_predictor}. Its mean, posterior variance, and sampling update are as follows:
\begin{equation}
\begin{aligned}
\boldsymbol{\mu}_{\theta}
&=
\frac{1}{\sqrt{\alpha_t}}
\left(
\mathbf{x}_t
-
\frac{\beta_t}{\sqrt{1-\overline{\alpha}_t}}
\widehat{\boldsymbol{\epsilon}}_t
\right),\\
\widetilde{\beta}_t
&=
\beta_t
\frac{1-\overline{\alpha}_{t-1}}
{1-\overline{\alpha}_t},\\
\mathbf{x}_{t-1}
&=
\boldsymbol{\mu}_{\theta}
+
\mathbb{I}[t>1]\sqrt{\widetilde{\beta}_t}\boldsymbol{\eta},
\qquad
\boldsymbol{\eta}\sim\mathcal{N}(\mathbf{0},\mathbf{I}).
\end{aligned}
\label{eq:reverse_ddpm}
\end{equation}
Here, $\overline{\alpha}_0=1$; inference uses the same schedule as training, begins from pure Gaussian noise, and applies a deterministic final step.

\subsection{Denoiser}
\label{sec:s4_denoiser}

The denoiser combines DiffWave-style residual and skip connections
\cite{kong2020diffwave} with bidirectional S4 layers \cite{gu2022s4},
following the SSSD/SSSD-ECG design
\cite{alcaraz2023sssd,alcaraz2023ecg}. It contains $L_r=30$ residual
blocks with residual width $C_r=64$, each comprising two bidirectional
S4 layers with state dimension $N=64$ per feature. Conditioning uses
a coarse gaze reference and identity--task feature-wise linear
modulation (FiLM).

The noisy velocity $\mathbf{x}_t$ and coarse reference
$\mathbf{x}_0^{\mathrm{co}}$ are concatenated into four channels and
mapped to $\mathbf{h}^{(0)}\in\mathbb{R}^{C_r\times S}$ through a
$1\times1$ convolution and ReLU. A projected task embedding
$g_{\mathrm{task}}\mathbf{W}_{\mathrm{task}}\mathbf{e}_{\mathrm{task}}$
is added across time, where the learned scalar $g_{\mathrm{task}}$ is
initialized to $0.1$. The diffusion step $t$ is encoded using
128 sinusoidal features with frequencies
$\{10^{4j/63}\}_{j=0}^{63}$ and transformed by two fully connected
layers with SiLU activations into
$\mathbf{d}_t\in\mathbb{R}^{512}$.

Each S4 layer represents temporal dynamics through a discretized
state-space model:
\begin{equation}
\begin{aligned}
\mathbf{s}_k
&=\overline{\mathbf{A}}\mathbf{s}_{k-1}
  +\overline{\mathbf{B}}\mathbf{u}_k,\\
\mathbf{y}_k
&=\mathbf{C}\mathbf{s}_k+\mathbf{D}\mathbf{u}_k,
\end{aligned}
\label{eq:discrete_ssm}
\end{equation}
where $\overline{\mathbf{A}}$ and $\overline{\mathbf{B}}$ are obtained
by bilinear discretization, $\mathbf{C}$ projects the latent state
to the output, and $\mathbf{D}$ provides a direct input--output
connection. With zero initial state, this recurrence is equivalent
to convolution with kernel taps
$\overline{\mathbf{K}}_j
=\mathbf{C}\overline{\mathbf{A}}^{j}\overline{\mathbf{B}}$,
$j=0,\ldots,S-1$. S4 uses HiPPO initialization and a structured
low-rank parameterization for efficient kernel computation
\cite{gu2020hippo,gu2022s4}; convolution is evaluated using FFTs.
Separate forward and backward kernels provide bidirectional context:
\begin{equation}
\begin{aligned}
\mathbf{y}_{\mathrm{bi}}
={}&\overline{\mathbf{K}}_{\rightarrow}*\mathbf{u}
+\operatorname{rev}\!\left[
\overline{\mathbf{K}}_{\leftarrow}
*\operatorname{rev}(\mathbf{u})\right] +\mathbf{D}\mathbf{u},
\end{aligned}
\label{eq:bidirectional_s4}
\end{equation}
where $*$ denotes convolution and $\operatorname{rev}$ reverses the
temporal dimension. Each S4 layer subsequently applies GELU,
feature mixing, dropout, residual addition, and layer normalization \cite{hendrycks2016gaussian}.

Within residual block $\ell$, a block-specific projection of
$\mathbf{d}_t$ is added to $\mathbf{h}^{(\ell-1)}$. A kernel-three
convolution expands the representation to $2C_r$ channels, and the
first S4 layer produces $\mathbf{y}_{\mathrm{pre}}^{(\ell)}$.
FiLM \cite{perez2018film} then incorporates the combined condition
$\mathbf{c}=[\mathbf{e}_{\mathrm{user}};
\mathbf{e}_{\mathrm{task}}]\in\mathbb{R}^{256}$.
A block-specific projection to $4C_r$ channels, temporal broadcasting,
and a zero-initialized $1\times1$ affine projection produce
$\boldsymbol{\gamma}^{(\ell)},\boldsymbol{\beta}^{(\ell)}
\in\mathbb{R}^{2C_r\times S}$. Modulation is applied as follows:
\begin{equation}
\widetilde{\mathbf{y}}^{(\ell)}
=
\mathbf{y}_{\mathrm{pre}}^{(\ell)}
\odot\left(1+\alpha_\ell\boldsymbol{\gamma}^{(\ell)}\right)
+\alpha_\ell\boldsymbol{\beta}^{(\ell)},
\label{eq:film_modulation}
\end{equation}
where $\alpha_\ell=\tanh(g_c^{(\ell)})$ is a learned scalar gain and
$\odot$ denotes element-wise multiplication.

The second S4 layer processes
$\widetilde{\mathbf{y}}^{(\ell)}$, and its output is split into gate
and filter features. The gated activation
$\mathbf{a}^{(\ell)}
=\sigma(\mathbf{y}_g^{(\ell)})
\odot\tanh(\mathbf{y}_f^{(\ell)})$,
where $\sigma$ is the sigmoid function, is projected through a
$1\times1$ convolution into residual and skip features
$\mathbf{r}^{(\ell)}$ and $\mathbf{k}^{(\ell)}$. Residual updates and skip aggregation use the following formulas:
\begin{equation}
\begin{aligned}
\mathbf{h}^{(\ell)}
&=\frac{\mathbf{h}^{(\ell-1)}+\mathbf{r}^{(\ell)}}{\sqrt{2}},\\
\mathbf{k}_{\mathrm{all}}
&=\frac{1}{\sqrt{L_r}}
\sum_{\ell=1}^{L_r}\mathbf{k}^{(\ell)}.
\end{aligned}
\label{eq:denoiser_aggregation}
\end{equation}
Finally, $\mathbf{k}_{\mathrm{all}}$ passes through a $1\times1$
convolution, ReLU, and a two-channel $1\times1$ convolution to predict
$\widehat{\boldsymbol{\epsilon}}_t\in\mathbb{R}^{2\times S}$.
The final convolution uses zero-initialized weights.

\subsection{Loss Functions}
\label{sec:loss_functions}

The denoiser is optimized using a diffusion noise-prediction objective augmented with identity, spectral, and displacement constraints. For the auxiliary losses, $\widehat{\mathbf{x}}_0$ is obtained by solving Equation~\eqref{eq:forward_diffusion} for $\mathbf{x}_0$ using the predicted noise, adding $10^{-8}$ to the denominator, and clipping the result to $[-1+10^{-3},1-10^{-3}]$. The reconstructed physical velocity $\widehat{\mathbf{v}}$ is then obtained by inverse sine normalization.

\subsubsection{Diffusion Noise-Prediction Loss}
The primary objective is the mean-squared error between the sampled
Gaussian noise and the denoiser output:
\begin{equation}
\mathcal{L}_{\mathrm{noise}}
=
\mathbb{E}_{b,c,s}
\left[
\left(
\epsilon_{b,c,s}
-
\widehat{\epsilon}_{t_b,b,c,s}
\right)^2
\right].
\label{eq:noise_loss}
\end{equation}

\subsubsection{Identity-Preservation Loss}
To retain subject-specific oculomotor characteristics, the real and
reconstructed physical velocities are passed through the frozen EKYT
encoder $\phi(\cdot)$. Their cosine distance defines
\begin{equation}
\mathcal{L}_{\mathrm{id}}
=
\mathbb{E}_{b}
\left[
1-
\frac{
\phi(\mathbf{v}_b)^{\mathsf{T}}
\phi(\widehat{\mathbf{v}}_b)
}{
\|\phi(\mathbf{v}_b)\|_2
\|\phi(\widehat{\mathbf{v}}_b)\|_2
}
\right].
\label{eq:identity_loss}
\end{equation}
Minimizing this term aligns the generated and real samples in the learned
biometric embedding space.

\subsubsection{Multi-Resolution Spectral Loss}
Temporal dynamics at several frequency resolutions are preserved using a magnitude-based short-time Fourier transform (STFT) loss on normalized velocities. With
$\mathcal{R}=\{64,128,256,512\}$,
\begin{equation}
\mathcal{L}_{\mathrm{spec}}
=
\frac{1}{|\mathcal{R}|}
\sum_{n\in\mathcal{R}}
\operatorname{mean}
\left(
\left|
\left|\operatorname{STFT}_{n}(\mathbf{x}_{0})\right|
-
\left|\operatorname{STFT}_{n}(\widehat{\mathbf{x}}_{0})\right|
\right|
\right).
\label{eq:spectral_loss}
\end{equation}
Here, the mean is taken over the minibatch, gaze-channel, frequency-bin,
and time-frame dimensions. Each transform uses a Hann window of length
$n$ and a hop length of $n/4$. A constant of $10^{-7}$ is added inside
the magnitude calculation for numerical stability.

\subsubsection{Integrated-Drift Loss}
Small velocity biases can accumulate into large position errors. We
therefore compare the final displacement obtained by integrating the real
and reconstructed physical velocities:
\begin{equation}
\mathcal{L}_{\mathrm{drift}}
=
\mathbb{E}_{b}
\left[
\frac{1}{2}
\left\|
\frac{1}{f_s}
\sum_{s=1}^{S}
\left(
\widehat{\mathbf{v}}_{b,s}-\mathbf{v}_{b,s}
\right)
\right\|_1
\right],
\label{eq:drift_loss}
\end{equation}
where $f_s$ is the sampling frequency. This term penalizes endpoint
displacement error and reduces long-term drift after velocity integration.

\subsubsection{Overall Training Objective}
The losses are combined using the weights employed in all experiments:
\begin{equation}
\mathcal{L}_{\mathrm{total}}
=
\mathcal{L}_{\mathrm{noise}}
+
0.5\mathcal{L}_{\mathrm{id}}
+
0.1\mathcal{L}_{\mathrm{spec}}
+
0.003\mathcal{L}_{\mathrm{drift}}.
\label{eq:total_loss}
\end{equation}

\section{Experiments}
\subsection{Dataset}
We trained and evaluated DiffSSMEye on GazeBase \cite{griffith2021gazebase}, a large-scale, publicly available, longitudinal eye-movement dataset. GazeBase contains 12,334 monocular recordings from 322 participants, collected over nine rounds spanning 37 months. Left-eye movements were recorded at 1,000 Hz using an EyeLink 1000 eye tracker. Each recording round comprised two sessions covering seven tasks: fixation (FXS), horizontal saccades (HSS), random oblique saccades (RAN), reading (TEX), two cinematic video-viewing tasks (VD1 and VD2), and gaze-driven gaming (BLG). This task diversity supports the evaluation of synthesis across different oculomotor behaviors. We adopted a subject-disjoint evaluation protocol, using recordings from 263 participants for training and 59 participants for testing. The test cohort comprised the participants recorded in Round~6. No recordings from these test participants were used to train DiffSSMEye, enabling evaluation on identities unseen during generative-model training. Spatial accuracy and precision were evaluated on HSS and RAN, using their corresponding target-position information to assess gaze behavior during target fixation. For recording-level evaluation, synthetic five-second segments were concatenated in their original temporal order and aligned with the corresponding real segments, enabling direct real--synthetic comparisons.

\subsection{Implementation and Training Details}

The DDPM was trained with $T=200$ diffusion steps and a linear noise schedule increasing $\beta_t$ from $10^{-4}$ to $2\times10^{-2}$. During training, diffusion timesteps were sampled uniformly from $\{1,\ldots,T\}$, and clean normalized velocities were corrupted with Gaussian noise according to the forward diffusion process. The conditional denoiser was optimized to predict the injected noise. AdamW \cite{loshchilov2017decoupled} was used with a learning rate of $2\times10^{-4}$, a batch size of 16, and weight decay of $0.01$. The gradient norm was clipped to a maximum of $1.0$. Training was configured for 200 epochs with a random seed of 1337.

The model was implemented in PyTorch~2.1.0 and trained on an NVIDIA RTX~A6000 GPU. Automatic mixed precision was enabled using \texttt{bfloat16} autocasting. TensorFloat-32 (TF32) computation was also enabled for CUDA matrix multiplication and cuDNN operations. The pre-trained EKYT identity encoder remained frozen throughout training. To support reproducibility, we will release the complete source code and trained model checkpoints.
\paragraph{Baseline Models.} We compare DiffSSMEye with SP-EyeGAN, a VAE baseline, and DiffEyeSyn. All models are trained and evaluated using the same subject-disjoint dataset split and evaluation protocol to ensure a consistent comparison.

\subsection{Evaluation Details}

\subsubsection{Spatial accuracy and precision.}
We evaluate spatial signal quality on the HSS and RAN tasks using each selected Round~6 subject--session recording. Gaze positions comprise horizontal and vertical coordinates $(x,y)$ expressed in degrees. Each sequence is divided into non-overlapping 80-sample bins, corresponding to 80~ms at 1000~Hz \cite{lohr2019evaluating}. Incomplete bins and bins containing changes in either target coordinate are excluded.

\textbf{Spatial accuracy} is computed as the mean angular separation between synthetic gaze and the target within each retained bin. For this calculation, each two-dimensional position is mapped to a normalized direction vector:
\begin{equation}
\mathbf{v}(x,y)
=
\frac{
\left[
\tan(\pi x/180),\,
\tan(\pi y/180),\,
1
\right]^{\mathsf{T}}
}{
\sqrt{
1+\tan^2(\pi x/180)+\tan^2(\pi y/180)
}
}.
\label{eq:gaze_direction_mapping}
\end{equation}
Let $\mathbf{v}_{u,b,k}$ denote the derived synthetic gaze direction for sample $k$ in bin $b$ of user $u$, and let $\mathbf{v}_{u,b}^{T}$ denote the direction associated with the constant target position. Bin-level spatial accuracy is
\begin{equation}
\theta_{u,b}
=
\frac{1}{n_b}
\sum_{k=1}^{n_b}
\frac{180}{\pi}
\arccos\!\left[
\operatorname{clip}
\left(
\mathbf{v}_{u,b,k}^{\mathsf{T}}\mathbf{v}_{u,b}^{T},
-1,1
\right)
\right],
\label{eq:spatial-accuracy}
\end{equation}
where $n_b=80$ and $\operatorname{clip}$ bounds the argument to
$[-1,1]$ for numerical stability.

\textbf{Spatial precision} is measured as the root-mean-square (RMS) angular displacement between consecutive samples. The implemented precision calculation uses spherical angular separation. For a given bin, let $a_k=\pi x_k/180$ and $b_k=\pi y_k/180$ denote the synthetic gaze coordinates in radians. The angular step between consecutive samples is
\begin{equation}
\begin{aligned}
c_k
&=
\cos b_k\cos b_{k-1}\cos(a_k-a_{k-1})
+\sin b_k\sin b_{k-1},\\
\delta_k
&=
\frac{180}{\pi}
\arccos\!\left[
\operatorname{clip}(c_k,-1,1)
\right].
\end{aligned}
\label{eq:sample_angular_displacement}
\end{equation}
The corresponding bin-level RMS precision is
\begin{equation}
\rho_{u,b}
=
\sqrt{
\frac{1}{n_b-1}
\sum_{k=2}^{n_b}\delta_k^2
}.
\label{eq:spatial-precision}
\end{equation}
Both metrics are expressed in degrees of visual angle. Lower $\theta_{u,b}$ indicates closer target alignment, whereas lower $\rho_{u,b}$ indicates less sample-to-sample angular variation.

We summarize the bin-level measurements using error and population percentiles \cite{aziz2024evaluation}. For each subject--session recording, $E50$ and $E95$ denote the 50th and 95th percentiles across its retained bins. For each task and metric, population percentiles are then computed across the corresponding recording-level error percentiles. Thus, $U50|E50$ is the median of recording-level median errors, and $U95|E95$ is the 95th percentile of recording-level 95th-percentile errors. Sessions are retained as separate observations in this aggregation.

\subsubsection{Real--synthetic embedding similarity.}
We assess biometric feature consistency using the frozen EKYT eye-movement encoder \cite{lohr2022ekyt}. 
For $N$ sequence pairs, let $\mathbf{z}_i^R$ and $\mathbf{z}_i^S$ denote the real and synthetic embeddings of pair $i$. Their mean cosine similarity is
\begin{equation}
\mathrm{CosSim}
=
\frac{1}{N}
\sum_{i=1}^{N}
\frac{
(\mathbf{z}_i^R)^{\mathsf{T}}\mathbf{z}_i^S
}{
\|\mathbf{z}_i^R\|_2
\|\mathbf{z}_i^S\|_2
}.
\label{eq:embedding-similarity}
\end{equation}
where higher values indicate closer alignment between real and synthetic signals in the learned embedding space. 

\section{Results}
\subsection{Spatial Accuracy}

Table~\ref{tab:spatial_accuracy} reports spatial accuracy for HSS and RAN, expressed as angular error relative to the target position in degrees of visual angle (dva). At $U50\mid E50$, DiffSSMEye achieves the lowest errors among the evaluated generative models: 3.20~dva for HSS and 3.22~dva for RAN, compared with 3.73 and 4.06~dva for DiffEyeSyn, 15.45 and 13.63~dva for SP-EyeGAN, and 18.65 and 13.50~dva for the VAE, respectively. The improvement over DiffEyeSyn is modest for both HSS and RAN, whereas both tasks show substantially lower median errors than the GAN and VAE baselines.

At $U95\mid E95$, DiffSSMEye achieves the lowest synthetic HSS error of 36.91~dva, compared with 38.61~dva for DiffEyeSyn, 52.73~dva for SP-EyeGAN, and 50.81~dva for the VAE. For RAN, DiffSSMEye obtains 27.19~dva, improving upon SP-EyeGAN (47.07~dva) and the VAE (34.75~dva), while DiffEyeSyn achieves a lower error of 26.77~dva. These results demonstrate improved median target alignment across both tasks, with task-dependent performance at higher percentiles. Nevertheless, all synthetic results remain above the corresponding real-gaze errors, indicating residual spatial discrepancies.

\begin{table}[h]
\centering
\caption{U\textbar{}E spatial accuracy for HSS and RAN. Ground truth denotes the real positional signal; $\downarrow$ indicates lower is better.}
\label{tab:spatial_accuracy}
\resizebox{0.35\columnwidth}{!}{%
\begin{tabular}{lcccc}
\toprule
\multirow{2}{*}{Model} & \multicolumn{2}{c}{U50\textbar{}E50 $\downarrow$} & \multicolumn{2}{c}{U95\textbar{}E95 $\downarrow$} \\
\cmidrule(lr){2-3} \cmidrule(lr){4-5}
 & HSS & RAN & HSS & RAN \\
\midrule
Ground truth & 0.89  & 0.92  & 30.85 & 21.80 \\
SP-EyeGAN    & 15.45 & 13.63 & 52.73 & 47.07 \\
VAE          & 18.65  & 13.50  & 50.81  & 34.75 \\
DiffEyeSyn   & 3.73  & 4.06  & 38.61 & 26.77 \\
Ours   & 3.20  & 3.22  & 36.91  & 27.19 \\
\bottomrule
\end{tabular}}
\end{table}

\subsection{Spatial Precision}

Table~\ref{tab:spatial_precision} reports spatial precision as the RMS angular displacement between consecutive gaze samples within the evaluated bins. Lower values indicate less sample-to-sample variability. At $U50\mid E50$, all synthetic models and real gaze yield 0.01~dva at the reported numerical precision. At $U95\mid E95$, the VAE obtains the lowest synthetic HSS value of 0.23~dva, DiffSSMEye and DiffEyeSyn tie at 0.35~dva, and SP-EyeGAN obtains 0.37~dva. For RAN, the VAE obtains the lowest value of 0.17~dva, followed by DiffSSMEye at 0.28~dva, DiffEyeSyn at 0.29~dva, and SP-EyeGAN at 0.41~dva.

All synthetic models exhibit lower $U95\mid E95$ values than real gaze, which reaches 0.86~dva for HSS and 1.14~dva for RAN. This pattern indicates reduced upper-percentile variability and may reflect smoothing of the generated trajectories. However, the evaluated bins are selected by target constancy rather than fixation classification, so their variability can include genuine eye movements as well as measurement noise. Lower RMS values therefore do not, by themselves, establish greater fidelity to real gaze. These results should be interpreted jointly with spatial accuracy and biometric embedding similarity when assessing synthetic signal quality.

\begin{table}[h]
\centering
\caption{U\textbar{}E spatial precision for HSS and RAN. Ground truth denotes the real positional signal; $\downarrow$ indicates lower is better.}
\label{tab:spatial_precision}
\resizebox{0.35\columnwidth}{!}{%
\begin{tabular}{lcccc}
\toprule
\multirow{2}{*}{Model} & \multicolumn{2}{c}{U50\textbar{}E50 $\downarrow$} & \multicolumn{2}{c}{U95\textbar{}E95 $\downarrow$} \\
\cmidrule(lr){2-3} \cmidrule(lr){4-5}
 & HSS & RAN & HSS & RAN \\
\midrule
Ground truth & 0.01 & 0.01 & 0.86 & 1.14 \\
SP-EyeGAN    & 0.01 & 0.01 & 0.37 & 0.41 \\
VAE          & 0.01 & 0.01 & 0.23 & 0.17 \\
DiffEyeSyn   & 0.01 & 0.01 & 0.35 & 0.29 \\
Ours   & 0.01 & 0.01 & 0.35 & 0.28 \\
\bottomrule
\end{tabular}}
\end{table}

\subsection{Cosine Similarity}
Table~\ref{tab:cosine_similarity} reports cosine similarity between real and synthetic EKYT embeddings, with higher values indicating closer alignment in the learned biometric feature space. DiffSSMEye achieves the highest mean similarity on six tasks and ties with DiffEyeSyn on HSS at the reported precision. Its mean similarities range from 0.91 for FXS and HSS to 0.95 for TEX, compared with 0.86--0.92 for DiffEyeSyn, 0.64--0.78 for the VAE, and 0.11--0.16 for SP-EyeGAN. The largest absolute improvement over DiffEyeSyn occurs for TEX (0.06), followed by FXS and RAN (0.05 each). Improvements are smaller for VD1 (0.04), VD2 (0.02), and BLG (0.01), indicating that the magnitude of the gain depends on the task. 
These results support improved consistency between real and synthetic signals in the EKYT biometric representation across multiple viewing behaviors.

\begin{table*}[h]
\centering
\caption{Cosine similarity between pre-trained EKYT embeddings of real and synthetic eye movement signals for the evaluated generative models.}
\label{tab:cosine_similarity}
\resizebox{0.8\textwidth}{!}{
\begin{tabular}{lccccccc}
\toprule
\multirow{2}{*}{Model} & \multicolumn{7}{c}{Task} \\
\cmidrule(lr){2-8}
 & BLG & FXS & HSS & RAN & TEX & VD1 & VD2 \\
\midrule
SP-EyeGAN  & $0.14 \pm 0.13$ & $0.11 \pm 0.15$ & $0.11 \pm 0.14$ & $0.13 \pm 0.14$ & $0.16 \pm 0.14$ & $0.13 \pm 0.15$ & $0.12 \pm 0.14$ \\
VAE         & $0.78 \pm 0.04$ & $0.64 \pm 0.06$ & $0.73 \pm 0.02$ & $0.71 \pm 0.03$ & $0.73 \pm 0.03$ & $0.73 \pm 0.04$ & $0.73 \pm 0.04$ \\
DiffEyeSyn & $0.92 \pm 0.04$ & $0.86 \pm 0.05$ & $0.91 \pm 0.03$ & $0.89 \pm 0.04$ & $0.89 \pm 0.03$ & $0.90 \pm 0.04$ & $0.90 \pm 0.03$ \\
Ours  & $0.93 \pm 0.09$ & $0.91 \pm 0.11$ & $0.91 \pm 0.12$ & $0.94 \pm 0.08$ & $0.95 \pm 0.07$ & $0.94 \pm 0.07$ & $0.92 \pm 0.10$ \\
\bottomrule
\end{tabular}
}
\end{table*}


\subsection{Qualitative Evaluations}
\begin{figure*}[t]
    \centering
    \includegraphics[width=\textwidth]{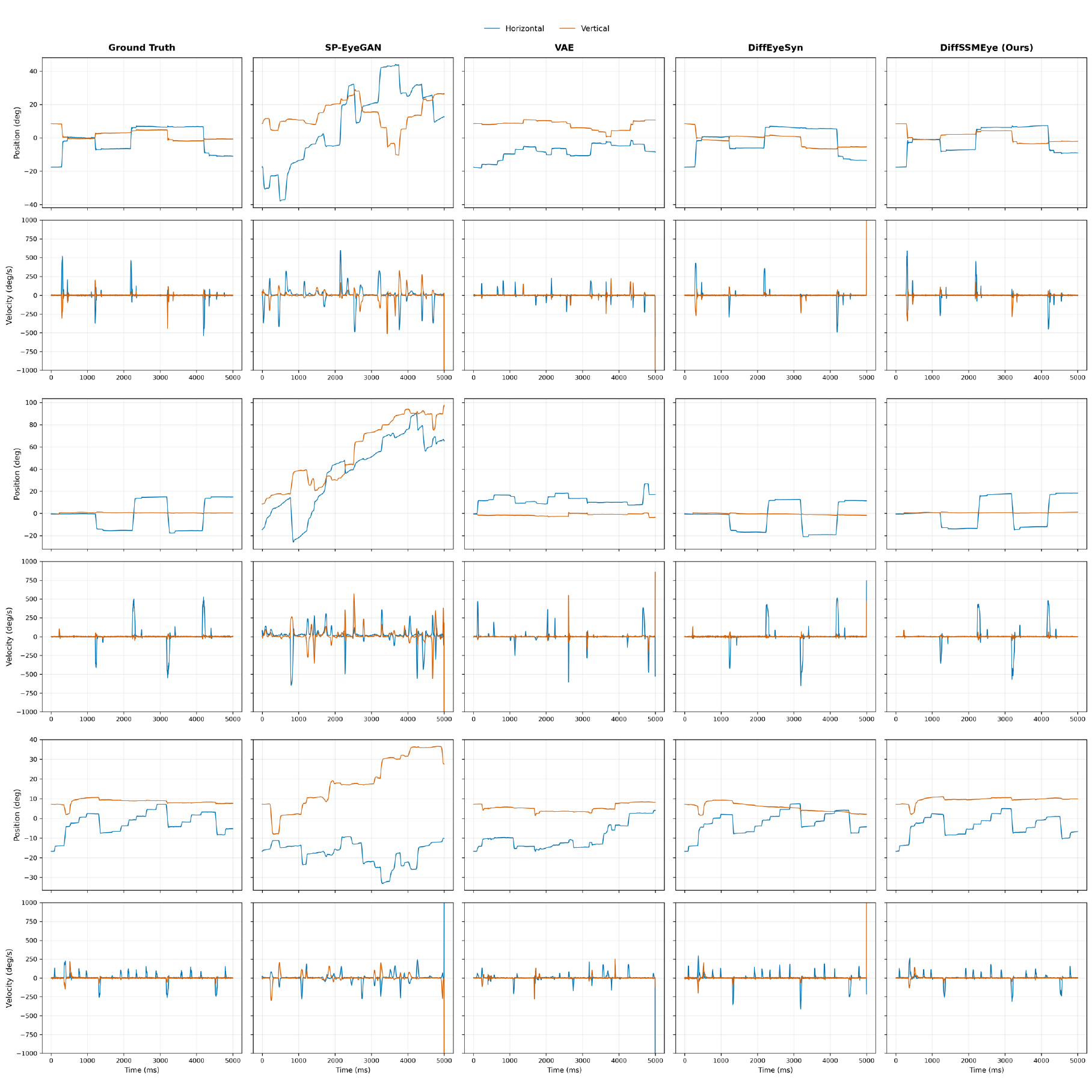}
    \caption{Qualitative real--synthetic gaze comparison for RAN, HSS, and TEX (top to bottom). Columns (left to right): Ground Truth, SP-EyeGAN, VAE, DiffEyeSyn, and DiffSSMEye. Each task occupies two rows: position above and velocity below. Subplots show horizontal (blue) and vertical (orange) components over 5~s; velocity axes span $[-1000,1000]$~deg/s for visual comparability.}
    \label{fig:qualitative_examples}
\end{figure*}

Figure~\ref{fig:qualitative_examples} compares real and synthetic gaze positions and velocities for the RAN, HSS, and TEX tasks. The examples reveal clear differences in how the models reproduce fixation plateaus, saccadic transitions, and velocity dynamics. SP-EyeGAN produces large or misplaced positional excursions and frequent velocity fluctuations that depart from the real trajectories. The VAE generates smoother signals, but several transitions are attenuated or misplaced, reducing their correspondence with the real movement patterns. DiffEyeSyn captures many major position changes more closely, although positional offsets and isolated velocity spikes remain visible, particularly near the sequence endpoints.

Meanwhile, DiffSSMEye preserves the main spatial and temporal patterns of the real signals across these examples. In RAN, it follows the major transitions in both gaze components while maintaining clear plateaus between movements. In HSS, it reproduces the alternating horizontal gaze positions and relatively stable vertical component. For TEX, it retains the denser sequence of smaller horizontal movements without the large drift observed in some baseline outputs. Its velocity peaks also align closely with the major transitions in the real signals, with less jitter than SP-EyeGAN and fewer extreme endpoint spikes than DiffEyeSyn. This combination of well-defined positional plateaus and localized velocity peaks makes the generated sequences visually consistent with the distinct movement patterns of each task.

\subsection{Statistical Analysis}
We conducted a subject-level statistical comparison of positional-signal reconstruction accuracy on held-out RAN recordings. All 118 recordings from 59 subjects were included, with two sessions per subject aggregated using the median session-level RMSE. Performance was quantified using two-dimensional vector RMSE over the first 5,000 samples, where lower values indicate closer agreement with the ground-truth trajectory. Non-finite values were replaced with zero in memory before analysis; no input files were modified. Median RMSE and median paired improvements were accompanied by 95\% percentile bootstrap confidence intervals based on 10,000 subject-level resamples. As summarized in Table~\ref{tab:ran_statistical_analysis}, DiffSSMEye achieved the lowest median RMSE (3.89) and showed positive median improvements over all baseline methods. Two-sided paired Wilcoxon signed-rank tests with Holm correction confirmed significant improvements over DiffEyeSyn ($p=0.006$), SP-EyeGAN ($p<0.001$), and VAE ($p<0.001$). These findings support improved RAN reconstruction accuracy for DiffSSMEye on the evaluated held-out samples.

\begin{table}[h]
\centering
\caption{Subject-level statistical comparison on held-out RAN positional signals. Lower RMSE is better; positive improvement favors DiffSSMEye.}
\label{tab:ran_statistical_analysis}

\makebox[\textwidth][c]{%
\resizebox{0.75\textwidth}{!}{%
\begin{tabular}{lccc}
\toprule
Model & Median RMSE [95\% CI] &
Improvement by DiffSSMEye [95\% CI] & Holm-adjusted $p$ \\
\midrule
DiffSSMEye (Ours) & 3.89 [3.14, 5.50] & --- & --- \\
DiffEyeSyn & 5.68 [4.00, 7.45] & 0.87 [0.20, 1.38] & 0.006 \\
SP-EyeGAN & 9.56 [9.19, 9.90] & 5.14 [3.86, 5.71] & $<0.001$ \\
VAE & 22.56 [20.06, 23.57] & 16.41 [14.33, 18.09] & $<0.001$ \\
\bottomrule
\end{tabular}%
}}

\begin{minipage}{0.7\textwidth}
\footnotesize
All results use the same cohort of 59 held-out subjects. Confidence intervals are 95\% subject-level bootstrap percentile intervals from 10,000 resamples and are not simultaneous intervals. P-values are from two-sided paired Wilcoxon signed-rank tests with Holm correction across three comparisons.
\end{minipage}
\end{table}

\subsection{Ablation Study}
\label{sec:ablation_study}
The ablation in Table~\ref{tab:conditioning_ablation} evaluates the contribution of each conditioning source while keeping the trained model, diffusion schedule, and sampling seed fixed. Without conditioning, DiffSSMEye obtains an EKYT cosine similarity of $0.87\pm0.06$. Reference and task conditioning individually provide modest improvements to $0.89\pm0.07$ and $0.88\pm0.07$, respectively, whereas identity conditioning produces the largest individual gain, reaching $0.92\pm0.07$. Combining reference and identity information further increases the similarity to $0.93\pm0.08$. The complete configuration achieves the highest similarity of $0.94\pm0.08$, representing an absolute improvement of $0.07$ over the unconditioned setting. These results indicate that identity information is the dominant conditioning signal, while reference and task cues provide complementary information that improves biometric alignment when jointly used. 

\begin{table*}[h]
\centering
\caption{Effect of conditioning inputs on real--synthetic embedding similarity. Checkmarks and dashes indicate present and absent conditions, respectively. Higher scores indicate closer embedding alignment.}
\label{tab:conditioning_ablation}
\resizebox{0.65\textwidth}{!}{%
\begin{tabular}{ccccc}
\toprule
\textbf{Configuration}
& \shortstack{\textbf{Reference}\\\textbf{conditioning}}
& \shortstack{\textbf{Identity}\\\textbf{conditioning}}
& \shortstack{\textbf{Task}\\\textbf{conditioning}}
& \shortstack{\textbf{EKYT cosine}\\\textbf{similarity} $\uparrow$} \\
\midrule
None                 & --            & --            & --            & 0.87 $\pm$ 0.06 \\
Reference only       & $\checkmark$  & --            & --            & 0.89 $\pm$ 0.07 \\
Task only            & --            & --            & $\checkmark$  & 0.88 $\pm$ 0.07 \\
Identity only        & --            & $\checkmark$  & --            & 0.92 $\pm$ 0.07 \\
Reference + Task     & $\checkmark$  & --            & $\checkmark$  & 0.92 $\pm$ 0.07 \\
Identity + Task      & --            & $\checkmark$  & $\checkmark$  & 0.92 $\pm$ 0.07 \\
Reference + Identity & $\checkmark$  & $\checkmark$  & --            & 0.93 $\pm$ 0.08 \\
All conditions       & $\checkmark$  & $\checkmark$  & $\checkmark$  & 0.94 $\pm$ 0.08 \\
\bottomrule
\end{tabular}%
}
\end{table*}

\section{Discussion}

This study examines whether high-frequency gaze synthesis can preserve spatial alignment and subject-specific biometric representations through a conditional diffusion model with bidirectional S4 layers. DiffSSMEye combines a coarse reference trajectory with identity and task conditioning, while auxiliary losses constrain biometric consistency, spectral structure, and integrated displacement. Across the evaluated generative models, it achieves the lowest median spatial errors for HSS and RAN, the highest real--synthetic EKYT similarity on six tasks, and a tie with DiffEyeSyn on HSS at the reported precision. These findings support the effectiveness of the complete framework, while also demonstrating that spatial accuracy, sample-to-sample variability, and biometric feature consistency capture distinct aspects of synthetic-gaze quality. 

The spatial-accuracy results show that improvements depend on both the task and the evaluated percentile. DiffSSMEye achieves $U50\mid E50$ errors of 3.20~dva for HSS and 3.22~dva for RAN, indicating stronger median target alignment than the evaluated baselines. At $U95\mid E95$, its HSS error of 36.91~dva is lower than those of the competing generators and close to the real-gaze value of 30.85~dva. However, DiffEyeSyn obtains a lower RAN error at this operating point, with 26.77~dva compared with 27.19~dva for DiffSSMEye. Thus, improved median alignment does not imply uniformly lower upper-percentile errors. Reporting both summaries reveals differences that would be obscured by a single aggregate score. Because the implemented aggregation treats subject--session recordings separately, these percentiles characterize the evaluated recording population rather than uniquely ranked users.

Spatial precision requires a complementary interpretation. All models yield a median RMS angular displacement of 0.01~dva at the reported numerical precision, whereas their upper-percentile values differ. VAE obtains the lowest synthetic HSS value of 0.23~dva, whereas both DiffSSMEye and DiffEyeSyn achieved 0.35~dva. Nevertheless, all generators exhibit substantially lower upper-percentile variability than real gaze. This pattern may reflect attenuation of rapid sample-to-sample changes, but it does not establish improved physiological fidelity. The evaluated intervals have constant target coordinates and are not explicitly restricted to fixations; consequently, their variability can include genuine oculomotor activity as well as measurement noise. A generator can therefore produce smoother trajectories while underrepresenting part of the real signal's temporal structure.

The embedding results further provide evidence of stronger biometric feature consistency across the evaluated tasks. DiffSSMEye achieves mean EKYT cosine similarities of 0.91--0.95. It exceeds all three baselines on six tasks and ties DiffEyeSyn on HSS at the reported precision. However, the same EKYT representation also supplies identity conditioning and identity-preservation supervision. These scores therefore measure alignment within a representation that directly guides generation.

The qualitative examples complement these measurements by illustrating differences in trajectory structure. In the displayed HSS, RAN, and TEX sequences, DiffSSMEye more closely reproduces fixation-like plateaus, major positional transitions, and localized velocity peaks. SP-EyeGAN exhibits larger positional excursions and velocity fluctuations in the illustrated examples, while the VAE attenuates some transitions. DiffEyeSyn reproduces several major trajectory features but also shows instances of channel-specific misalignment and saturated velocity peaks. These observations are consistent with the goal of combining temporal modeling with identity, spectral, and displacement constraints. 

\section{Limitations and Future Work}

Evaluation is limited to five-second GazeBase segments; generalization to other datasets, sampling rates, acquisition conditions, and longer sequences remains untested. Because EKYT is used for conditioning, supervision, and evaluation, biometric assessment is not independent, and the current metrics do not establish downstream authentication utility or complete physiological fidelity. Future work should examine cross-dataset and long-sequence generation, independent biometric encoders, authentication protocols, event-level oculomotor characteristics, and synthetic-data augmentation.

\section{Conclusion}
We presented DiffSSMEye, a conditional diffusion framework with bidirectional S4 layers for generating five-second, user-specific gaze-velocity sequences at 1000~Hz. The framework combines coarse-reference, identity, and task conditioning with auxiliary objectives for biometric consistency, spectral structure, and integrated displacement. In subject-disjoint evaluation on GazeBase, DiffSSMEye achieves the highest mean real--synthetic EKYT cosine similarity across on six tasks, ranging from 0.91 to 0.95. It also achieves the lowest median spatial errors among the evaluated generators for HSS and RAN, at 3.20 and 3.22~dva, respectively. The results demonstrate improved median target alignment and consistency of the biometric features captured by EKYT. Together, these findings support DiffSSMEye as an effective framework for high-frequency gaze synthesis that incorporates both task context and subject-specific biometric information.


\section*{Use of Generative AI} A large language model (Astra, OpenAI) was used in preparing this manuscript beyond copy-editing, in accordance with ACM policy on generative AI. It assisted with drafting and revising prose, including portions of the Related Work and Discussion, and with restructuring the manuscript across multiple revisions. It was not used to collect data, conduct the study, implement the measurement pipeline, or generate experimental results. All reported results and numerical values originate from the authors' own experiments and analysis code. The authors critically reviewed and revised all AI-assisted content and take full responsibility for the manuscript's content, analysis, and conclusions.

\section*{Privacy and Ethics Statement}
This study uses anonymized, publicly available eye-tracking data to train generative models, posing minimal societal risk. No personal data was used, and we encourage responsible application to avoid potential misuse. Our work supports ethical, privacy-conscious advancements in eye-tracking research.

\bibliographystyle{unsrtnat}
\bibliography{references}  

@String{Computing = "Computing" }

@String{Computer = "{IEEE} Computer" }

@article{griffith2021gazebase,
  title={GazeBase, a large-scale, multi-stimulus, longitudinal eye movement dataset},
  author={Griffith, Henry and Lohr, Dillon and Abdulin, Evgeny and Komogortsev, Oleg},
  journal={Scientific Data},
  volume={8},
  number={1},
  pages={184},
  year={2021},
  publisher={Nature Publishing Group UK London}
}

@article{haller2022eye,
  title={Eye-tracking based classification of Mandarin Chinese readers with and without dyslexia using neural sequence models},
  author={Haller, Patrick and S{\"a}uberli, Andreas and Kiener, Sarah Elisabeth and Pan, Jinger and Yan, Ming and J{\"a}ger, Lena},
  journal={arXiv preprint arXiv:2210.09819},
  year={2022}
}

@article{thanarajan2023eye,
  title={Eye-Tracking Based Autism Spectrum Disorder Diagnosis Using Chaotic Butterfly Optimization with Deep Learning Model.},
  author={Thanarajan, Tamilvizhi and Alotaibi, Youseef and Rajendran, Surendran and Nagappan, Krishnaraj},
  journal={Computers, Materials \& Continua},
  volume={76},
  number={2},
  year={2023}
}

@article{lohr2022ekyt,
  title={Eye know you too: Toward viable end-to-end eye movement biometrics for user authentication},
  author={Lohr, Dillon and Komogortsev, Oleg V},
  journal={IEEE Transactions on Information Forensics and Security},
  volume={17},
  pages={3151--3164},
  year={2022},
  publisher={IEEE}
}

@article{patney2016towards,
  title={Towards foveated rendering for gaze-tracked virtual reality},
  author={Patney, Anjul and Salvi, Marco and Kim, Joohwan and Kaplanyan, Anton and Wyman, Chris and Benty, Nir and Luebke, David and Lefohn, Aaron},
  journal={ACM Transactions On Graphics (TOG)},
  volume={35},
  number={6},
  pages={1--12},
  year={2016},
  publisher={ACM New York, NY, USA}
}

@article{menges2019improving,
  title={Improving user experience of eye tracking-based interaction: Introspecting and adapting interfaces},
  author={Menges, Raphael and Kumar, Chandan and Staab, Steffen},
  journal={ACM Transactions on Computer-Human Interaction (TOCHI)},
  volume={26},
  number={6},
  pages={1--46},
  year={2019},
  publisher={ACM New York, NY, USA}
}

@inproceedings{hutt2021breaking,
  title={Breaking out of the lab: Mitigating mind wandering with gaze-based attention-aware technology in classrooms},
  author={Hutt, Stephen and Krasich, Kristina and R. Brockmole, James and K. D'Mello, Sidney},
  booktitle={Proceedings of the 2021 CHI conference on human factors in computing systems},
  pages={1--14},
  year={2021}
}

@inproceedings{singh2011eye,
  title={Eye tracking based driver fatigue monitoring and warning system},
  author={Singh, Hardeep and Bhatia, Jagjit Singh and Kaur, Jasbir},
  booktitle={India International Conference on Power Electronics 2010 (IICPE2010)},
  pages={1--6},
  year={2011},
  organization={IEEE}
}

@inproceedings{steil2019privacy,
  title={Privacy-aware eye tracking using differential privacy},
  author={Steil, Julian and Hagestedt, Inken and Huang, Michael Xuelin and Bulling, Andreas},
  booktitle={Proceedings of the 11th ACM Symposium on Eye Tracking Research \& Applications},
  pages={1--9},
  year={2019}
}

@inproceedings{lan2022eyesyn,
  title={Eyesyn: Psychology-inspired eye movement synthesis for gaze-based activity recognition},
  author={Lan, Guohao and Scargill, Tim and Gorlatova, Maria},
  booktitle={2022 21st ACM/IEEE international conference on information processing in sensor networks (IPSN)},
  pages={233--246},
  year={2022},
  organization={IEEE}
}

@inproceedings{duchowski2015modeling,
  title={Modeling physiologically plausible eye rotations},
  author={Duchowski, Andrew T and J{\"o}rg, Sophie},
  booktitle={Proceedings of computer graphics international},
  year={2015}
}

@inproceedings{assens2018pathgan,
  title={PathGAN: Visual scanpath prediction with generative adversarial networks},
  author={Assens, Marc and Giro-i-Nieto, Xavier and McGuinness, Kevin and O'Connor, Noel E},
  booktitle={Proceedings of the European Conference on Computer Vision (ECCV) Workshops},
  pages={0--0},
  year={2018}
}

@inproceedings{simon2016automatic,
  title={Automatic scanpath generation with deep recurrent neural networks},
  author={Simon, Daniel and Sridharan, Srinivas and Sah, Shagan and Ptucha, Raymond and Kanan, Chris and Bailey, Reynold},
  booktitle={Proceedings of the ACM symposium on applied perception},
  pages={130--130},
  year={2016}
}

@article{fuhl2018eye_Santini,
  title={Eye movement simulation and detector creation to reduce laborious parameter adjustments},
  author={Fuhl, Wolfgang and Santini, Thiago and Kuebler, Thomas and Castner, Nora and Rosenstiel, Wolfgang and Kasneci, Enkelejda},
  journal={arXiv preprint arXiv:1804.00970},
  year={2018}
}

@article{fuhl2018eye_Kasneci,
  title={Eye movement velocity and gaze data generator for evaluation, robustness testing and assess of eye tracking software and visualization tools},
  author={Fuhl, Wolfgang and Kasneci, Enkelejda},
  journal={arXiv preprint arXiv:1808.09296},
  year={2018}
}

@inproceedings{duchowski2016eye,
  title={Eye movement synthesis},
  author={Duchowski, Andrew T and J{\"o}rg, Sophie and Allen, Tyler N and Giannopoulos, Ioannis and Krejtz, Krzysztof},
  booktitle={Proceedings of the ninth biennial ACM symposium on eye tracking research \& applications},
  pages={147--154},
  year={2016}
}

@inproceedings{campbell2014saliency,
  title={Saliency-based bayesian modeling of dynamic viewing of static scenes},
  author={Campbell, Daniel J and Chang, Joseph and Chawarska, Katarzyna and Shic, Frederick},
  booktitle={Proceedings of the symposium on eye tracking research and applications},
  pages={51--58},
  year={2014}
}

@article{yeo2012eyecatch,
  title={Eyecatch: Simulating visuomotor coordination for object interception},
  author={Yeo, Sang Hoon and Lesmana, Martin and Neog, Debanga R and Pai, Dinesh K},
  journal={ACM Transactions on Graphics (TOG)},
  volume={31},
  number={4},
  pages={1--10},
  year={2012},
  publisher={ACM New York, NY, USA}
}

@inproceedings{ma2009natural,
  title={Natural eye motion synthesis by modeling gaze-head coupling},
  author={Ma, Xiaohan and Deng, Zhigang},
  booktitle={2009 IEEE Virtual Reality Conference},
  pages={143--150},
  year={2009},
  organization={IEEE}
}

@inproceedings{lee2002eyes,
  title={Eyes alive},
  author={Lee, Sooha Park and Badler, Jeremy B and Badler, Norman I},
  booktitle={Proceedings of the 29th annual conference on Computer graphics and interactive techniques},
  pages={637--644},
  year={2002}
}

@article{welch1995introduction,
  title={An introduction to the Kalman filter},
  author={Welch, Greg and Bishop, Gary and others},
  year={1995},
  publisher={Chapel Hill, NC, USA}
}

@article{kong2020diffwave,
  title={Diffwave: A versatile diffusion model for audio synthesis},
  author={Kong, Zhifeng and Ping, Wei and Huang, Jiaji and Zhao, Kexin and Catanzaro, Bryan},
  journal={arXiv preprint arXiv:2009.09761},
  year={2020}
}

@article{goodfellow2020generative,
  title={Generative adversarial networks},
  author={Goodfellow, Ian and Pouget-Abadie, Jean and Mirza, Mehdi and Xu, Bing and Warde-Farley, David and Ozair, Sherjil and Courville, Aaron and Bengio, Yoshua},
  journal={Communications of the ACM},
  volume={63},
  number={11},
  pages={139--144},
  year={2020},
  publisher={ACM New York, NY, USA}
}

@article{jiao2025diffgaze,
  title={DiffGaze: A Diffusion Model for Modelling Fine-grained Human Gaze Behaviour on 360° Images},
  author={Jiao, Chuhan and Wang, Yao and Zhang, Guanhua and B{\^a}ce, Mihai and Hu, Zhiming and Bulling, Andreas},
  journal={ACM Transactions on Interactive Intelligent Systems},
  year={2025},
  publisher={ACM New York, NY}
}

@article{de2022next,
  title={Next-generation deep learning based on simulators and synthetic data},
  author={De Melo, Celso M and Torralba, Antonio and Guibas, Leonidas and DiCarlo, James and Chellappa, Rama and Hodgins, Jessica},
  journal={Trends in cognitive sciences},
  volume={26},
  number={2},
  pages={174--187},
  year={2022},
  publisher={Elsevier}
}

@inproceedings{fuhl2022hpcgen,
  title={Hpcgen: Hierarchical k-means clustering and level based principal components for scan path genaration},
  author={Fuhl, Wolfgang and Kasneci, Enkelejda},
  booktitle={2022 Symposium on Eye Tracking Research and Applications},
  pages={1--7},
  year={2022}
}

@inproceedings{fuhl2021fully,
  title={Fully convolutional neural networks for raw eye tracking data segmentation, generation, and reconstruction},
  author={Fuhl, Wolfgang and Rong, Yao and Kasneci, Enkelejda},
  booktitle={2020 25th International Conference on Pattern Recognition (ICPR)},
  pages={142--149},
  year={2021},
  organization={IEEE}
}

@inproceedings{prasse2023sp,
  title={Sp-eyegan: Generating synthetic eye movement data with generative adversarial networks},
  author={Prasse, Paul and Reich, David Robert and Makowski, Silvia and Ahn, Seoyoung and Scheffer, Tobias and J{\"a}ger, Lena A},
  booktitle={Proceedings of the 2023 symposium on eye tracking research and applications},
  pages={1--9},
  year={2023}
}

@article{ho2020denoising,
  title={Denoising diffusion probabilistic models},
  author={Ho, Jonathan and Jain, Ajay and Abbeel, Pieter},
  journal={Advances in neural information processing systems},
  volume={33},
  pages={6840--6851},
  year={2020}
}

@article{jiao2024diffeyesyn,
  title={DiffEyeSyn: Diffusion-based User-specific Eye Movement Synthesis},
  author={Jiao, Chuhan and Zhang, Guanhua and Cho, Yeonjoo and Hu, Zhiming and Bulling, Andreas},
  journal={arXiv preprint arXiv:2409.01240},
  year={2024}
}

@inproceedings{aziz2024evaluation,
  title={Evaluation of eye tracking signal quality for virtual reality applications: A case study in the meta quest pro},
  author={Aziz, Samantha and Lohr, Dillon J and Friedman, Lee and Komogortsev, Oleg},
  booktitle={Proceedings of the 2024 Symposium on Eye Tracking Research and Applications},
  pages={1--8},
  year={2024}
}

@article{savitzky1964smoothing,
  title={Smoothing and differentiation of data by simplified least squares procedures.},
  author={Savitzky, Abraham and Golay, Marcel JE},
  journal={Analytical chemistry},
  volume={36},
  number={8},
  pages={1627--1639},
  year={1964},
  publisher={ACS Publications}
}

@article{lohr2019evaluating,
  title={Evaluating the data quality of eye tracking signals from a virtual reality system: Case study using smi's eye-tracking htc vive},
  author={Lohr, Dillon J and Friedman, Lee and Komogortsev, Oleg V},
  journal={arXiv preprint arXiv:1912.02083},
  year={2019}
}

@article{hasan2025quantitative,
  title={Quantitative and Qualitative Comparison of Generative Models for Subject-Specific Gaze Synthesis: Diffusion vs GAN},
  author={Hasan, Kamrul and Katrychuk, Dmytro and Raju, Mehedi Hasan and Komogortsev, Oleg V},
  journal={arXiv preprint arXiv:2511.09867},
  year={2025}
}

@article{loshchilov2017decoupled,
  title={Decoupled weight decay regularization},
  author={Loshchilov, Ilya and Hutter, Frank},
  journal={arXiv preprint arXiv:1711.05101},
  year={2017}
}

@article{raju2021determining,
  title={Determining which sine wave frequencies correspond to signal and which correspond to noise in eye-tracking time-series},
  author={Raju, Mehedi H and Friedman, Lee and Bouman, Troy M and Komogortsev, Oleg V},
  journal={Journal of Eye Movement Research},
  volume={14},
  number={3},
  pages={16},
  year={2021},
  publisher={Bern Open Publishing}
}

@article{hendrycks2016gaussian,
  title={Gaussian error linear units (gelus)},
  author={Hendrycks, Dan and Gimpel, Kevin},
  journal={arXiv preprint arXiv:1606.08415},
  year={2016}
}

@article{rayner2009eye,
  title={Eye movements and attention in reading, scene perception, and visual search},
  author={Rayner, Keith},
  journal={The quarterly journal of experimental psychology},
  volume={62},
  number={8},
  pages={1457--1506},
  year={2009},
  publisher={Taylor \& Francis}
}

@misc{eyelink1000plus,
  author       = {{SR Research}},
  title        = {{EyeLink 1000 Plus Technical Specifications}},
  year         = {2026},
  howpublished = {\url{https://www.sr-research.com/eyelink-1000-plus-technical-specifications/}},
  note         = {Accessed: 2026-10-11}
}

@misc{tobiiprospectrum,
  author       = {{Tobii Pro}},
  title        = {{Tobii Pro Spectrum}},
  year         = {2026},
  howpublished = {\url{https://www.tobii.com/products/eye-trackers/screen-based/tobii-pro-spectrum}},
  note         = {Accessed: 2026-10-11}
}

@article{merzon2022eye,
  title={Eye movement behavior in a real-world virtual reality task reveals ADHD in children},
  author={Merzon, Liya and Pettersson, Kati and Aronen, Eeva T and Huhdanp{\"a}{\"a}, Hanna and Seesj{\"a}rvi, Erik and Henriksson, Linda and MacInnes, W Joseph and Mannerkoski, Minna and Macaluso, Emiliano and Salmi, Juha},
  journal={Scientific reports},
  volume={12},
  number={1},
  pages={20308},
  year={2022},
  publisher={Nature Publishing Group UK London}
}

@article{gu2022s4,
  title={Efficiently modeling long sequences with structured state spaces},
  author={Gu, Albert and Goel, Karan and R{\'e}, Christopher},
  journal={arXiv preprint arXiv:2111.00396},
  year={2021}
}

@article{gu2020hippo,
  title={Hippo: Recurrent memory with optimal polynomial projections},
  author={Gu, Albert and Dao, Tri and Ermon, Stefano and Rudra, Atri and R{\'e}, Christopher},
  journal={Advances in neural information processing systems},
  volume={33},
  pages={1474--1487},
  year={2020}
}

@article{alcaraz2023sssd,
  title={Diffusion-based time series imputation and forecasting with structured state space models},
  author={Alcaraz, Juan Miguel Lopez and Strodthoff, Nils},
  journal={arXiv preprint arXiv:2208.09399},
  year={2022}
}

@article{alcaraz2023ecg,
  title={Diffusion-based conditional ECG generation with structured state space models},
  author={Alcaraz, Juan Miguel Lopez and Strodthoff, Nils},
  journal={Computers in biology and medicine},
  volume={163},
  pages={107115},
  year={2023},
  publisher={Elsevier}
}

@inproceedings{perez2018film,
  title={Film: Visual reasoning with a general conditioning layer},
  author={Perez, Ethan and Strub, Florian and De Vries, Harm and Dumoulin, Vincent and Courville, Aaron},
  booktitle={Proceedings of the AAAI conference on artificial intelligence},
  volume={32},
  number={1},
  year={2018}
}






\end{document}